\documentclass[aps,prl,twocolumn,groupedaddress,showpacs]{revtex4}
\usepackage{graphicx}

\pdfoutput=1

\begin{document}


\title{Entropic Spectral Broadening in Carbon Nanotube Resonators}


\author{Arthur W. Barnard$^1$}
\author{Vera Sazonova$^2$}
\author{Arend M. van der Zande$^2$}
\author{Paul L. McEuen$^{2,3}$}
\affiliation{$^1$Applied and Engineering Physics, $^2$Laboratory of Atomic and Solid State Physics, and $^3$Kavli Institute at Cornell for Nanoscale Science, Cornell University, Ithaca, New York 14853, USA}

\affiliation{}


\date{\today}

\begin{abstract}

We simulated the behavior of suspended carbon-nanotube (CNT) resonators over a broad range of temperatures to address the unexplained spectral broadening and frequency shifts seen in experiments. We find that thermal fluctuations induce strong coupling between resonance modes. This effect leads to spectral fluctuations which readily account for the experimentally observed quality factors $Q\sim100$ at $300\ K$. Using a mean field approach to describe entropic fluctuations we analytically calculate $Q$ and frequency shifts in tensioned and buckled CNTs and find excellent agreement with simulations.

\end{abstract}


\maketitle


Carbon nanotube (CNT) resonators show great promise for fundamental science and nanomechanical applications owing to their high stiffness, low mass, electrical detectability, and defect-free structure. Although experimental work has shown CNT resonators to be highly tunable\cite{Sazonova:2004zm} as well as functional as RF transceivers\cite{doi:10.1021/nl0721113} and atomic mass detectors\cite{JensenK.:2008rt,Lassagne:2008rt}, these resonators have consistently exhibited much poorer than expected properties\cite{EichlerA.:2011kl, Sazonova:2004zm}. Specifically, the quality factor $Q$, the key parameter measuring the degree to which an oscillating mode is decoupled from its environment, is typically less than $Q\sim100$ at room temperature. 

These low $Q$s are not consistent with known dissipation mechanisms seen in other nanomechanical resonator systems \cite{Cleland:2002ef}. Analytical phonon-phonon scattering studies establish a theoretical upper bound on $Q$ in CNTs that is well above experimental values, with $Q\lesssim50,000$ at $300\ K$  \cite{PhysRevB.74.165420,PhysRevB.81.233409}. Molecular-dynamics simulations of short CNTs ($L\sim\ 50nm$) show interesting behaviors in cantilevered and free CNT segments \cite{PhysRevLett.92.015901,PhysRevB.81.125436,PhysRevLett.93.185501, Greaney:2009ys} due to anharmonic atomic potentials, and those that make an explicit determination of quality factor in thermal equilibrium \cite{PhysRevB.81.125436,PhysRevLett.93.185501} give $Q\sim1,000$  at room temperature. Typically, however, experiments are performed on CNTs with $L\gtrsim1\ \mu m$, and the anharmonic elastic effects that dominate in short length CNTs do not contribute as significantly at this longer length scale. 

 \begin{figure}
 \includegraphics[width=3.3in]{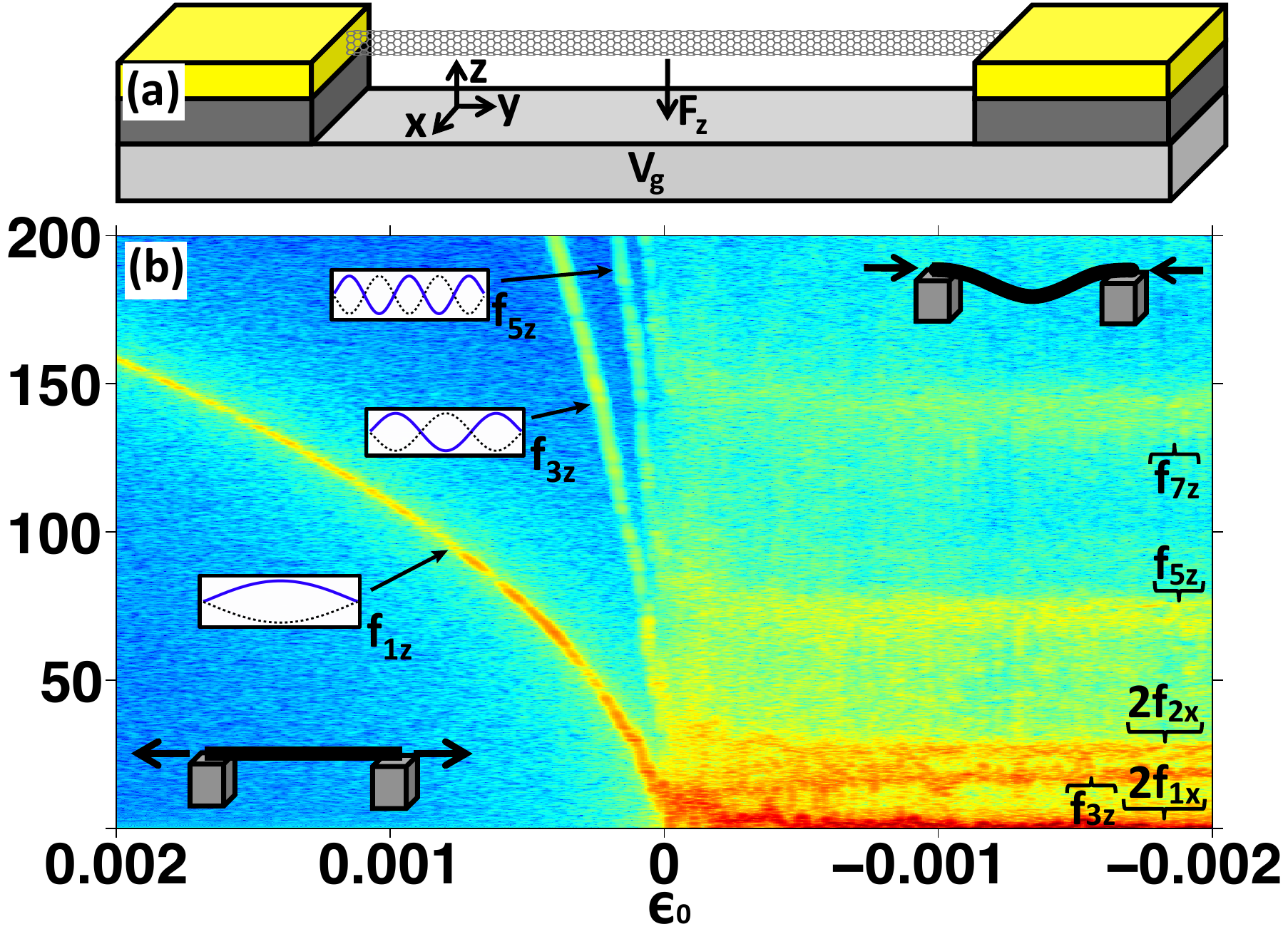}
  \vspace{-10pt}
  \hspace{-10pt}
 \caption{ a) Schematic of CNT resonator geometry. CNT is suspended between two electrodes (in yellow) and has a controllable downward force $F_z$ electrostatically applied by a voltage to the underlying substrate. A small $F_z$ is applied in b) to keep the buckled tube oriented vertically. b)Simulated spectral power density of the z-motion of the nanotube as a function of strain $\epsilon_0$ at $T=100\ K$. The tube is taken from tensile strain on the left to compressive strain on the right, as illustrated by the insets. Eigenmode shapes are inset, as are labels for the eigenfrequencies.   \label{}}
 \vspace{-10pt}
 \end{figure}

A key difference between CNT resonators and other nanomechanical resonators is their nanometer-scale cross section\cite{4877395}, yielding a very small bending rigidity for flexural modes and hence large amplitude fluctuations in thermal equilibrium. The effects of these thermal fluctuations on the physics of 1D elastic objects has a long and storied history. A key length-scale that characterizes their properties is the persistence length $\ell_p=\frac{\kappa}{k_BT}$, where $\kappa$ is the bending rigidity. Short 1D structures ($L\ll \ell_p$) such as microfabricated NEMS are dominated by elastic effects and thus behave like rigid rods, while long 1D structures ($L\gg \ell_p$) such as organic polymers are dominated by configurational entropic effects caused by thermal forces and thus take on random 3D configurations. Micron-scale CNTs behave as semi-flexible polymers ($L\lesssim \ell_p$)\cite{Fakhri25082009}, where the bending energy and configurational entropy contribute comparably to the total free energy. The interplay of these effects in semi-flexible polymers is an active area of research. In overdamped, aqueous environments, these effects are now well understood \cite{PhysRevE.73.031906}, but high-amplitude fluctuations on the properties of nanoscale resonators has not been explored. 

 In this Letter, we report simulations of the thermally driven dynamics of CNT resonators over temperatures from $3\ K$ to $300\ K$ and in the limits of both compressive and tensile strain. By simulating purely continuum elastic behavior, we isolate the specific influence of entropic forces on measured quantities, including both quality factor and thermally-induced frequency shifts. Quantum mechanical effects are neglected, as the thermal occupation of all modes considered are in the classical limit. We find that the fluctuations of many resonance modes strongly influence the quality factor of CNT resonators, an effect we call entropic spectral broadening.  We find that this entropic spectral broadening is in good agreement with observed quality factors in experiment. 
 
 CNTs were modeled as 1D elastic objects, with bending rigidity $\kappa=\pi\frac{C d^3}{8}$ and extensional rigidity $K=\pi C d$, where $C\approx 345\ J/m^2$ \cite{PhysRevB.64.235406} is the 2D elastic modulus  of graphene and $d$ is the tube diameter. Each CNT was discretized into 100 masses joined by axial and torsional springs. For a given set of tube dimensions, boundary conditions, and externally applied forces, the equilibrium geometry was computed via a relaxation method and zero-temperature eigenmodes were computed by diagonalizing the force constant matrix \cite{Ustunel:2005ai}. 
 
 	In order to study dynamics at non-zero temperature, a finite time-difference calculation was employed. Thermal equilibrium was reached by coupling the nanotube to an external heat-bath via a generalized Langevin equation, applying a stochastic white noise force with external damping. Dynamics at equilibrium were then simulated with this Langevin equation using the $4^{th}$ order Runge-Kutta method. Adiabatic conditions were also studied using Stoermer's rule\cite{Press:1992:NRC:148286}. The model's validity was confirmed by its ability to properly exhibit equipartition of energies for each eigenmode. 

 	Throughout this Letter, data are shown from simulations of a typical experimental case of a CNT with $L=3\ \mu m$ and $d=2\ nm$. We investigate resonance properties of this CNT at different zero-temperature strains  $\epsilon_0$, temperatures $T$, and externally applied forces $F_z$ by quasi-statically sweeping at most one parameter and measuring the power spectral density of the mean $z$-displacement (defined in fig 1a)  of the nanotube:
$
S_z \left( f \right)=\lim_{t_m \rightarrow \infty} E\left [ \frac{\left | \mathcal{F}\right( z_{t_m}\left(t\right)\left) \right |^2}{t_m}\right]
$                                                                          
where $t_m$ is the finite time over which the Fourier transform is taken and $E$ denotes an ensemble average. The mean z-displacement was the chosen parameter to analyze as it is typically measured in experiments \cite{VeraThesis}. From $S_z(f)$, we measure thermal frequency shifts $\Delta f\equiv f-f(T=0K)$ and quality factor $Q\equiv\frac{\delta f}{f}$. This definition of $Q$ is employed as the linewidth $\delta f$ is directly measured in experiments.

We first study the qualitative behavior of this CNT as a function of strain at $100\ K$. In fig 1b, we plot a logarithmic color map of $S_z \left( f \right)$ with red as the highest amplitude, and blue the lowest. In the tensioned limit, to the left of fig 1b, the eigenmodes, labeled by their modeshape, tune like a tensioned string, giving a frequency: $f_n\approx\frac{n}{2L}\sqrt{\frac{N}{\mu}} $ where $\mu$ is the linear mass density and $N=K\epsilon$ is the axial tension. Modes without mean z-displacements, including the even in-plane (z-direction) eigenmodes and all out-of-plane (x-direction) eigenmodes, are not visible. Thus, only modes with $n=1,3,5...$ are seen.
	
	With negative strain, the nanotube undergoes a Euler buckling transition, as illustrated in the upper right corner of fig 1b. With the resulting built-in slack, the CNT can bend without stretching leading the spectral lines to no longer tune significantly with strain. The linewidths, however do tune with strain. Also, there are emergent spectral lines corresponding to the motion of out-of-plane modes, as labeled to the right of fig. 1b. We analyze these temperature-dependent linewidths in the tensioned case and buckled case separately and discuss the different nonlinear coupling mechanisms \cite{Nayfeh:1974fe,Conley:2008ys} that apply to each.

  \begin{figure}
 \includegraphics[width=3.3in]{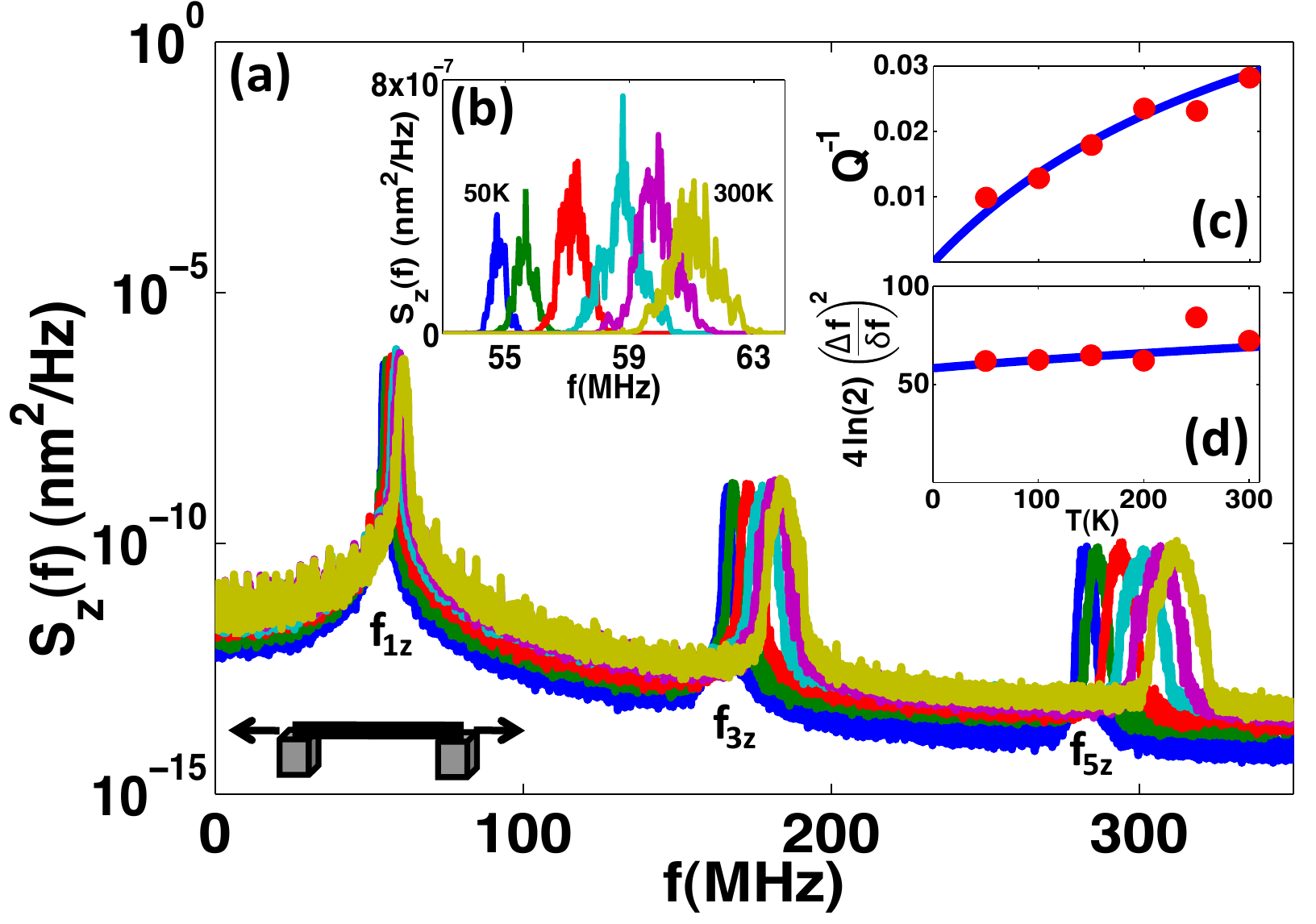}
 \vspace{-10pt}
 \caption{ a) Power spectral density of a tensioned nanotube with $\epsilon_0=2\times 10^{-4}$. Six temperatures are plotted from 50-300 $K$ spaced evenly in $T$. b) is a linear plot of $S_z(f)$ for $f_{1z}$ showing that the frequency shift $\Delta f$ and linewidth $\delta f$ both scale linearly with $T$ c) inverse quality factor of $f_{1z}$ vs. temperature. d) dimensionless fixed ratio of $\Delta f$  to $\delta f$  that corresponds to the number of independent fluctuating modes contributing to spectral broadening and frequency shifts. In c) and d) data are circles, and theory is a line. The line in c) corresponds to Eqn. 2 and in d) $n_f$ in Eqn. 1\label{}}
 \vspace{-10pt}
 \end{figure}
 
  	We begin with the tensioned case. We select a fixed tensile zero-temperature clamping condition $\epsilon_0=2\times 10^{-4}$, and simulate the nanotube motion at 6 temperatures, incrementing by 50 $K$ from 50 $K$ to 300 $K$ and plot in fig. 2a the spectral density over a frequency range spanning up to the $n=5$ mode. The frequencies of the modes shift linearly with $T$ and, as shown in Fig. 2c. The peaks also broaden with $Q^{-1}\sim T $.  At 300K, $Q\sim40$. 
		 
   	The quality factor and frequency shifts can be understood to arise from the change in length of the CNT caused by thermal fluctuation in each eigenmode.  The $n^{th}$ eigenmode is given by $u_{na}(y,t)=a_{n}(t)\xi_{na}(\frac{y}{L})$, where $\xi_{na}$ is the dimensionless modeshape with unit RMS displacement, $a_{n}$ is a time varying amplitude function, and both the amplitude $a$ and index $a$ refers to either the $x$ or $z$ direction. The tube elongates by the length $\beta_{na}\frac{a_n^2}{2L}$ where $\beta_{na}\equiv\int_0^1{\xi'_{na}(x)^2dx}$ for each independent eigenmode.
This leads to the strain being modified as: $
\left\langle \epsilon\right\rangle\approx \epsilon _0 +\sum_{n,a} \beta_{na}\frac{\left\langle a_{n}^2\right\rangle}{2L^2}$ 
where brackets denote a time average over the period of oscillation \cite{PhysRevLett.105.117205}.  Assuming that $\left\langle a_{n}\right\rangle$ obey Boltzmann statistics in thermal equilibrium and that $a_{n}$ fluctuate incoherently, the first and second moment of the strain shift are $\overline{\Delta \epsilon} = \sum_{na} \beta_{na} \frac{\overline{a_{n}^2}}{2L^2} $ and  $\sigma_\epsilon^2=\sum_{n,a}\beta_{na}^2 \left(\frac{\overline{a_{n}^4}}{4L^4}-\frac{\overline{a_{n}^2}^2}{4L^4}\right)$ respectively, with thermal averages $\overline{a_{n}^2}=\frac{k_BT}{k_{na}}$ and $\overline{a_{n}^4}=2\left(\frac{k_BT}{k_{na}}\right)^2$. Here $k_{na}$ is the effective spring constant: $k_{na}\approx \beta_{na}\frac{K\epsilon}{L}+\alpha_{na}\frac{\kappa}{L^3}$ where $\alpha_{na}\equiv\int_0^1{\xi''_{na}(x)^2dx}$ parameterizes the bending associated with the $n^{th}$ eigenmode. In the case of a tensioned string, $\beta_{na}\approx n^2\pi^2$ and $\alpha_{na}\approx n^4\pi^4$.
 
In the high tension limit, a large number of the resonance modes contribute significantly to length fluctuations. High frequency modes, however do not contribute significantly, as their high bending stiffness limits their thermal amplitudes. Approximating the infinite series as an integral gives: 
\begin{equation}
\overline{\Delta \epsilon}=\frac{L}{2n_f\ell_p}\ ;\ \ n_f=\sqrt{\frac{NL^2}{\kappa}}
\end{equation}
where the value of $\overline{\Delta \epsilon}$ is solved self-consistently and $n_f$ can be interpreted as the number of independent degrees of freedom with significant fluctuation amplitudes. $\overline{\Delta \epsilon}$ is a length-dependent thermal expansion parameter that is distinct from the intrinsic thermal expansion simulated for CNTs \cite{PhysRevLett.92.015901}. From Eqn. 1, $\overline{\Delta \epsilon}$ leads to ${\Delta f}\sim T$ . 

 Next, we calculate the strain variance:
$
\sigma_\epsilon^2= \frac{L^2}{8n_f^3\ell_p^2}	
$. 
 This leads to the prediction that ${\delta f}\sim T$.  The resulting dimensionless quantity $4\ln(2)\left(\frac{\Delta f}{\delta f}\right)^2$, plotted in fig. 2d is a direct measurement of $n_f$.  Furthermore, the predicted ${\delta f}$ constitutes entropic spectral broadening and gives a unique prediction for $Q$:
\begin{equation}
Q^{-1}=\frac{\sqrt{\ln2}}{2} \frac{L}{ n_f^{\frac{3}{2}}\epsilon \ell_p}.                 
\end{equation}
 The dominant $T$-dependence comes from $\ell_p^{-1}\sim T$, but both  $n_f$ and $\epsilon$ have weak $T$-dependence as a result of entropic stretching. The quantity $n_f^2\ell_p$ is the persistence length for a tensioned beam and parameterizes how far fluctuations drive the CNT out of equilibrium. The $n_f^\frac{3}{2}$ dependence then can be understood to arise from an additional linewidth broadening $\sim n_f^\frac{1}{2}$ due to averaging over $n_f$ uncorrelated degrees of freedom. Eqn. 2 accurately describes the numerical results, shown in fig. 2c, and shows that entropic spectral broadening can account for the experimentally observed $Q\sim100$ at room temperature. 

 \begin{figure}
 \includegraphics[width=3.3in]{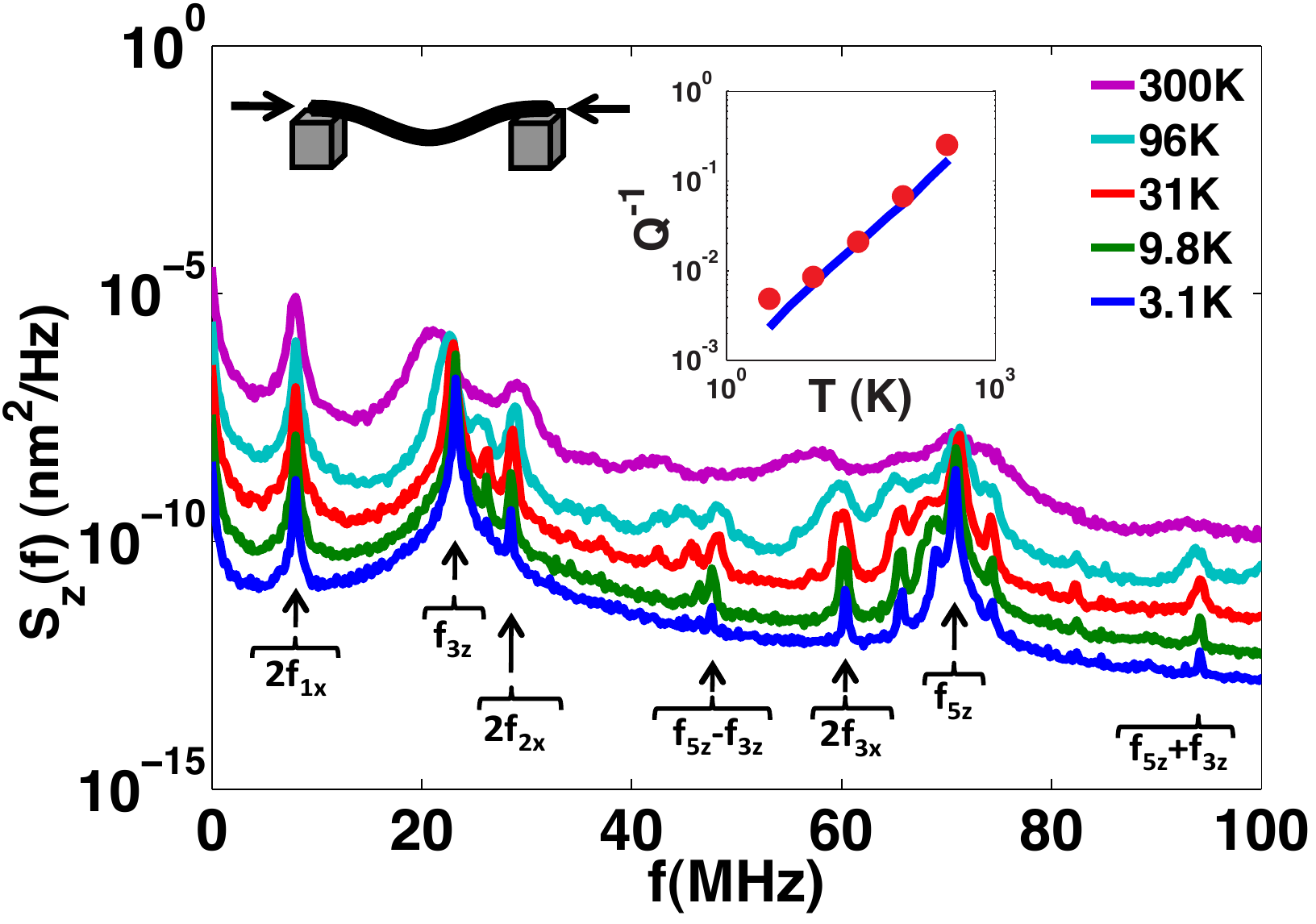}
  \vspace{-10pt}
 \caption{
 Spectral power density of buckled nanotube with $\epsilon_0=-4\times 10^{-3}$ with $F_z=0.6\ pN$ at logarithmically spaced temperatures. The built-in intrinsic damping dominates the apparent linewidth at the lowest temperature. Inset is a plot of $Q^{-1}$ of the $f_{3z}$ mode. Simulation data are shown as red dots, and theoretical predictions from Eqn. 3 are plotted as a blue line. Labels beneath each spectral line correspond to their expected origin. \label{}}
 \vspace{-10pt}
 \end{figure}
 
  Next, we study a nanotube under compressive strain, picking  $\epsilon_0=-4\times 10^{-3}$. Applying a small downward force $F_z=0.6\ pN$ that mimics the force applied by the gate in experiments, simulations were performed at logarithmically spaced temperatures from 3.1 $K$ to 300 $K$, as shown in fig. 3.  In contrast to the tensioned regime, there is complex structure in the spectra, with many spectral features growing non-linearly with $T$. Linear theory predicts that $f_{3z}$ and $f_{5z}$ would be the only visible spectral lines in fig. 3. The other emergent modes can be identified as either oscillations of out-of-plane modes producing $z$-displacement at twice their natural frequency ($2f_{1x}$, $ 2f_{2x}$, and $2f_{3x}$) and mixes of in-plane modes ($f_{5z}-f_{3z}$ and $f_{5z}+f_{3z}$).  Focusing on the lowest observed in-plane mode ($f_{3z}$), we observe $Q^{-1}\sim T$ (fig. 2 inset) and a decrease in frequency with increasing temperature. From this we extract $Q\sim5$ for this mode at $300\ K$. 

Further insight is gained by smoothly tuning frequencies by quasi-statically varying the magnitude of the electrostatic force $F_z=\frac{1}{2} C' V_g^2$, as shown in fig. 4a, where $V_g$ corresponds to the electrostatic voltage applied to the gate illustrated in fig 1b and $C'=\frac{dC}{dz}$ is the derivative of the tube-gate capacitance assuming a $400\ nm$ gap. The spectral lines tune with force and  the frequency-doubled out-of-plane spectral lines cut across the in-plane spectral lines. At $0\ K$, linear theory predicts there to be no coupling. However, at the intersections, there are observable avoided crossings, even at $10\ K$ as seen in Fig. 4a. This behavior is consistent with experiments. To study this, we fix $F_z$ at a crossing indicated by the dotted line in fig. 4a, quasi-statically sweep the temperature, and measure $S_z(f)$ (shown in the inset of fig. 4a). Here, we observe a sub-linear $T$-dependence of the splitting at the avoided crossing. At $300\ K$ the frequency splitting is nearly $1/5^{th}$ of the resonance frequency, indicating a strong coupling strength induced entirely by thermal fluctuations.
 
We build an analytical model for this behavior by first modeling the CNT as an inextensible object. In this framework, when an out-of-plane mode has finite amplitude, the CNT must move up in the $z$-direction to preserve length. To quantify the amount that the tube moves in $z$, we enforce the constraint that the differential length $dL=\sqrt{\beta_{1z}\left|\epsilon\right|}dz_1-2\sum_n \frac{\beta_{1x} x_{n}}{L}dx_{n}$ is zero, where the equilibrium deflection of the tube is $z_1\xi_{1z}(\frac{y}{L})$.  $\xi_{1z}$ is a linear combination of in-plane modes, requiring to $2^{nd}$ order that $z_{nNL}=z_{n}+\frac{\eta_n}{L}\sum_m \beta_{mx} x_{m}^2$ where $\eta_n\equiv \frac{\left\langle \xi_{1z}|\xi_{nz} \right \rangle}{ \left \langle \xi_{nz}|\xi_{nz}\right\rangle}\frac{2}{\sqrt{\beta_{1z}\left | \epsilon  \right |}}$ in which the brackets are an inner product defined by $\left\langle a(x)|b(x)\right\rangle =\int_0^1a(x)b(x)dx$ \cite{PhysRevLett.105.117205}. 

 \begin{figure}
 \includegraphics[width=3.3in]{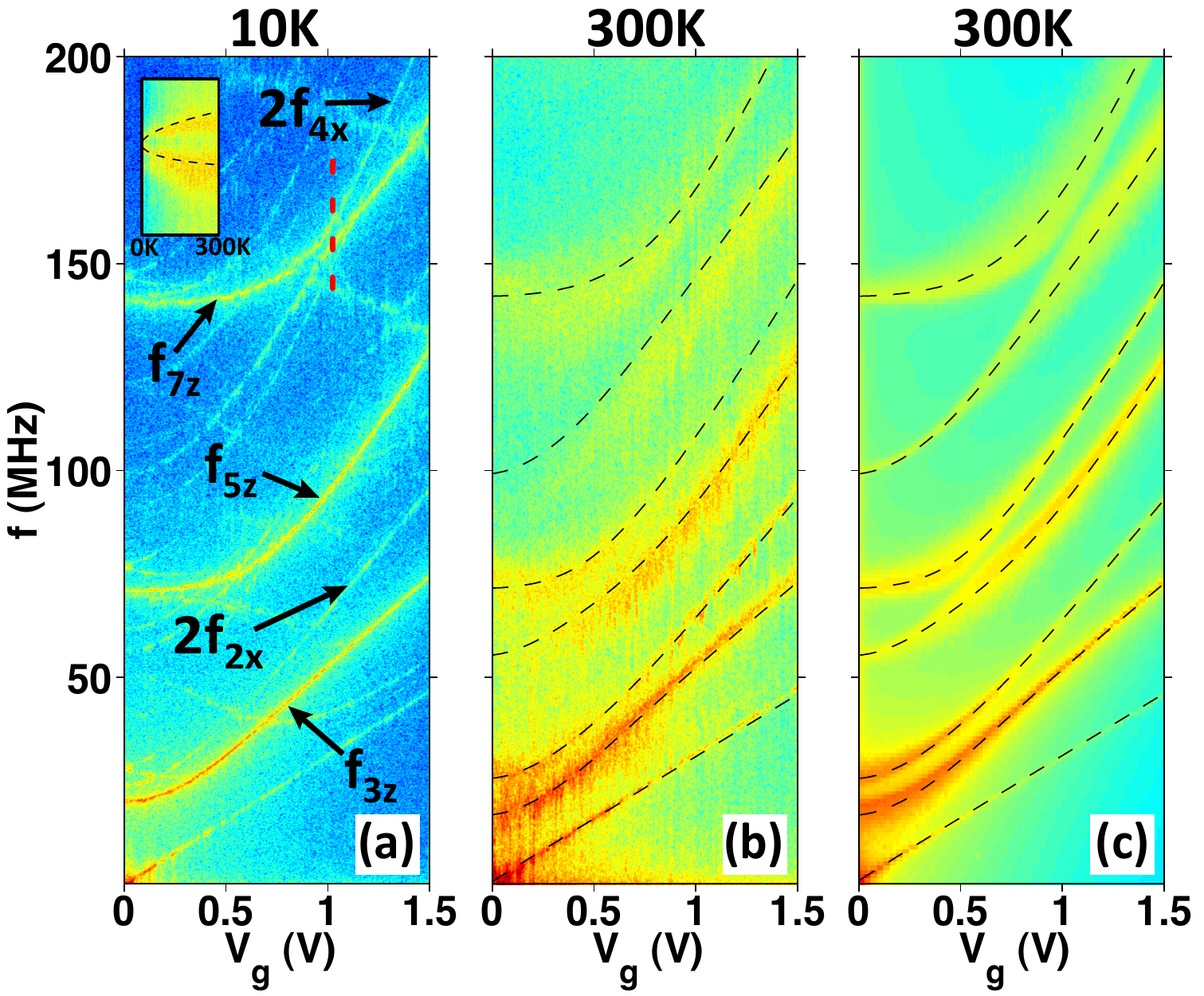}
  \vspace{-10pt}
 \caption{a) color-map of spectral noise density at 10 $K$ of the mean z-displacement of a buckled nanotube resonator as a function of gate voltage assuming tube is 400nm above the gate. Modes are labeled as defined in fig. 3. T-dependence from 0 $K$ to 300 $K$ of the apparent avoided-crossing at $150\ MHz$ is shown in the inset at a fixed $V_g$ indicated by the red dotted line b) 300 $K$ data of same force sweep in a). c) plots the theoretical spectral noise density, based on the generalized version of Eqn. 3 described in the text, for the same conditions as in b). The dashed lines in b) and c) correspond to the theoretical central frequencies from the generalized version of Eqn. 3.   \label{}}
 \vspace{-10pt}
 \end{figure}

 	Focusing on the interaction of one in-plane and one out-of-plane mode, we change the out-of-plane variable to $z_{op}=\frac{\eta}{L}x_{op}^2$ and apply time-dependent perturbation theory \cite{Nayfeh} to generate a coupled set of linearized equations:
\begin{equation}
\left[\begin{array}{cc}
k_{ip}&-\alpha(T)k_{ip}\\ 
-\alpha(T)k_{ip} & 4k_{op}+\alpha^2(T)k_{ip}
\end{array}\right] \left[\begin{array}{c} z_{ip}\\ z_{op} \end{array}\right] =m\left[\begin{array}{c} \ddot{z}_{ip}\\ \ddot{z}_{op} \end{array}\right]
\end{equation}
 where $\alpha(T)\equiv\frac{2\eta\sqrt{\left \langle x_{op}^2\right \rangle}}{L}$ and $\left \langle x_{op}^2\right \rangle$ is time-averaged over the period of oscillation.
 
   Using $\left \langle x_{op}^2\right \rangle=\frac{k_BT}{k_{op}}$ gives $\alpha(T)\sim\sqrt{T}$, which when substituted into Eqn. 3 predicts average frequency shifts. At an avoided crossing, $k_{ip}\approx4k_{op}$ gives the prediction that $\Delta f\approx \frac{1}{2}\alpha(T)f_{ip}$. This matches the simulated $T$-dependence, as shown by the theoretical prediction that is overlaid on top of the data in the inset of fig. 4a. 
	
	By extending Eqn. 3 to the full interaction matrix, populating each mode according to Boltzmann statistics, and weighting the resulting frequency probability distribution by the squared-amplitude distribution we are able to generate a theoretical $S_z(f)$ map, which is shown in fig. 4c. It compares well with the simulated results at $300\ K$ shown in fig. 4b. In addition, the linewidths in the theoretical $S_z(f)$  give an accurate prediction of the simulated $Q$s, as shown in the inset of fig. 3. The specific degree of coupling is geometry dependent, and thus not fully analytically generalizable, but Eqn. 3 predicts for the lowest in-plane mode $f_{3z}$ that:
	\begin{equation}
	Q^{-1}\approx 0.05 \frac{L}{|\epsilon| \ell_p}.             
	\end{equation}
	  Eqn. 4 explains the strain dependence of the spectral fluctuations seen in the right half of fig. 1a: $Q$ improves with increased buckling as the geometric coupling between modes decreases. Furthermore, due to its coupling with higher frequency modes, the lowest in-plane mode is predicted to decrease frequency with increasing temperature at low $V_g$, which is frequently observed in experiment \cite{Lassagne:2008rt,VeraThesis}. 

These results show that entropic spectral broadening dominates the behavior of CNT resonators over a broad range of temperatures, and appears to be the main cause of temperature-dependent quality factors measured in both tensioned and untensioned resonators. To date there is limited experimental data characterizing $Q$ over a broad temperature range, but thus far, data remain at or below our theoretical upper bound\cite{VeraThesis,HuAattel:2009rp,EichlerA.:2011kl}. This work implies a fundamental limit on $Q$ in high aspect ratio resonators at finite temperatures. Tailoring the geometry of these systems is necessary to mitigate thermally-induced mode-coupling and thus improve $Q$.

\begin{acknowledgments}
We thank Tomas Arias, Johannes Lischner, Richard Rand, James Sethna, Mohammad  Younis, and Alan Zehnder for fruitful discussions. This work was supported by NSF Grant\#0928552 and by the Cornell Center for Materials Research (CCMR)with funding from IGERT:  A Graduate Traineeship in Nanoscale Control of Surfaces and Interfaces, (DGE-0654193) and made use of the Research Computing facility of the CCMR with support from the National Science Foundation Materials Research Science and Engineering Centers (MRSEC) program (DMR 0520404). 
 \end{acknowledgments}


\end{document}